\documentclass[twocolumn,showpacs,preprintnumbers,amsmath,amssymb]{revtex4-1}
\usepackage{graphicx}% Include figure files
\usepackage{dcolumn}% Align table columns on decimal point
\usepackage{bm}% bold math
\begin{document}

%\preprint{APS/123-QED}

\title{Period tripling causes rotating spirals in agitated wet granular layers}% Force line breaks with \\

\author{Kai Huang}
\email{kai.huang@uni-bayreuth.de}
\author{Ingo Rehberg}
\affiliation{Experimentalphysik V, Universit\"at Bayreuth, 95440 Bayreuth, Germany}

\date{\today}% It is always \today, today,

\begin{abstract}

Pattern formation of a thin layer of vertically agitated wet granular matter is investigated experimentally. Rotating spirals with three arms, which correspond to the kinks between regions with different colliding phases, are the dominating pattern. This preferred number of arms corresponds to period tripling of the agitated granular layer, unlike predominantly subharmonic Faraday crispations in dry granular matter. The chirality of the spatiotemporal pattern corresponds to the rotation direction of the spirals.

\end{abstract}

% PACS

%Granular systems, classical mechanics of, 45.70.-n
%Pattern formation in granular systems, 45.70.Qj  .+r in fluid dynamics
%Dynamical systems nonlinear, 05.45.-a
%Wave fronts, 42.15.Dp

%Granular flow
%classical mechanics of discrete systems, 45.70.Mg
%complex fluids, 47.57.Gc

\maketitle

%We consider the manuscript 'Period tripling causes rotating spirals in agitated wet granular layers' is being of general interest and thus suitable for publication in Physical Review Letter: It reveals a general understanding of spiral formation as a consequence of the spatiotemporal chirality stemming from a period tripling bifurcation.

From the formation of galaxies to hurricanes, from the structure of a seashell to the magnetic ordering on the atomic scale\cite{Ferriani08}, spirals are ubiquitous in nature \cite{[See e.g. ]spiralzoom}. During the past decades, various spiral forming systems have been extensively investigated, e.g. Rayleigh-B\'{e}nard convection \cite{Morris93, *Plapp98}, the Belousov-Zhabotinsky reaction \cite{Winfree72, *Tyson88, *Steinbock93, *Vanag01}, and cardiac tissue \cite{Davidenko92, *Fenton08}. Moreover, spacecraft observations of the Saturn's rings -- granular matter consisting of icy particles -- reveal that the perturbations arising from the gravitational field of its moons propagate in the form of spiral waves \cite{Rosen88, *Cassini10}, which have been modeled in terms of a hydrodynamical description \cite{Spahn06}. Despite this success, a continuum description of both dense and dilute granular systems \cite{Goldhirsch03, *AransonRMPl, *Pouliquen06, *Liu07l}, including wet granular systems \cite{Bocquet02, *Herminghaus05} is still far from complete.

In this Letter, we present rotating spiral kink waves formed in agitated wet granular layers. We demonstrate that the dominating three armed spiral is a manifestation of the period tripling in the system, and that the spatiotemporal chirality associated with period tripling drives the rotation of spiral arms. In contrast to those in dry granular systems \cite{Bruyn01l}, the arms are kinks separating regions colliding with the container at different phases. Understanding this particular spiral forming system could not only provide a step towards a continuum description of dry and wet granular matter, but also broaden our knowledge about the more general spiral forming systems mentioned above.

\begin{figure}
\includegraphics[width = 0.40\textwidth]{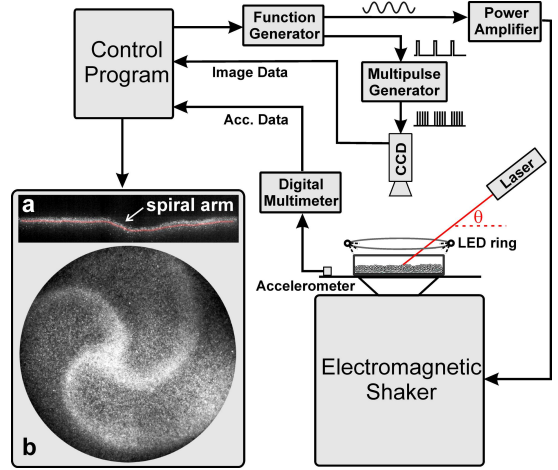}% Here is how to import EPS art
\caption{\label{setup}(color online) The experimental setup with sample images of a spiral pattern. Vertical or horizontal information of the patterns formed is obtained by a high speed CCD camera mounted above the sample illuminated by a laser sheet or a low angle LED ring correspondingly. The camera is triggered by a multi-pulse generator. Snapshot (a) is captured with the laser sheet illumination. The red half transparent line is a surface profile determined by an image processing procedure. The spiral arm corresponds to a kink between regions with different heights. Snapshot (b) corresponds to the low angle illumination (averaged over three synchronized images to enhance the contrast). }
\end{figure}

\begin{figure}
\includegraphics[width = 0.40\textwidth]{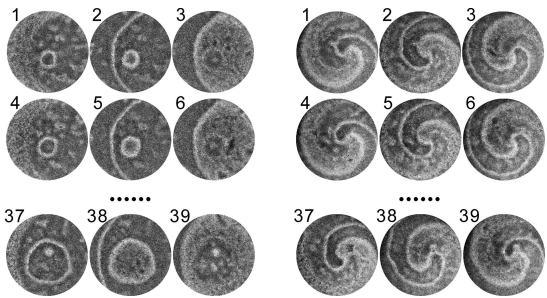}% Here is how to import EPS art
\caption{\label{snapshots} Snapshots of migrating fronts (left panel) and a rotating spiral (right panel) captured at a fixed phase of continuous vibration cycles (indicated as numbers) showing the period tripling behavior. Parameters: $f=80$\,Hz, $W=1.6\%$, $m=96$\,g and $\Gamma=36$(left), $37$(right)}
\end{figure}

Fig.~\ref{setup} shows a sketch of the experimental setup. The granular sample is prepared by adding purified water (Laborstar TWF-DI) into glass spheres (SiLiBeads S) with a diameter $d=0.78$\,mm and 10\% polydispersity. They were cleaned subsequently with ethanol, acetone and purified water, and dried in an oven before use. The liquid content $W=V_{\rm w}/V_{\rm g}$, where $V_{\rm w}$ is the volume of the water and $V_{\rm g}$ that of the glass beads, is kept within a few percent. A cylindrical polycarbonate container with an inner radius $R=8$\,cm and a height $1.05$\,cm is mounted on the electromagnetic shaker (Tira TV50350). The amount of sample (mass $m$) added corresponds to a height of a few particle diameters. The sinusoidal shaking frequency $f$ and amplitude of the shaker are controlled by the function generator (Agilent FG33220). The nondimensional acceleration $\Gamma=4\pi^2 f^2 A/g$, with vibration amplitude $A$ and gravitational acceleration $g$, is obtained by an accelerometer (Dyson 3035B2).

A high speed camera (IDT MotionScope M3) mounted above the container is used to capture the patterns: the LED ring illumination provides the horizontal motion of the patterns, while the surface profile is obtained via the laser profilometry method \cite{Raton95s}. The laser sheet illuminates a line on the sample surface with an angle $\theta=38^{\circ}$. The camera is triggered by a synchronized multi-pulse generator to capture images at fixed phases of each vibration cycle.

Typical patterns observed are rotating spirals, which coexist with migrating fronts  occasionally generated from the side wall or nucleated within the bulk (see Fig.~\ref{snapshots}). One striking feature arising from the snapshots is that both patterns show a period tripling behavior: similar patterns appear at every third vibration cycle, in contrast to the subharmonic patterns commonly seen in agitated dry granular matter \cite{Faraday1830, Melo95l}. The migrating fronts tend to merge with one of the spiral arms and vanish, because both are of the same origin: they are kinks separating regions of different phase shift within the period-3 vibration, as will be demonstrated below. The rotation period for the spiral shown in the right panel is $\approx 1.8$\,s, much larger than that for the driving $1/80$\,s.

\begin{figure}
\includegraphics[width = 0.4\textwidth]{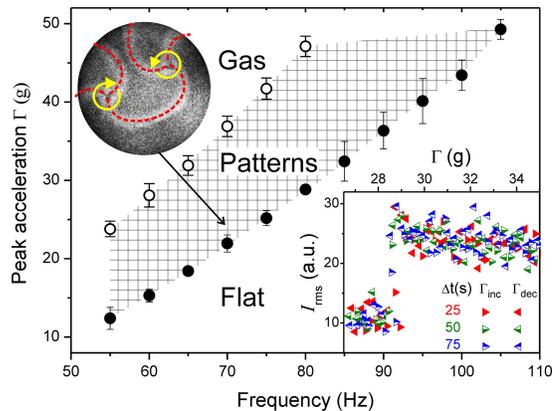}% Here is how to import EPS art
\caption{\label{pd} (color online) Phase diagram for the patterns (shaded with squares) measured by varying $\Gamma$ while keeping the vibration frequency for $m=129$\,g and $W=1.6\%$. The region with $\Gamma>50$ is unexplored. The snapshot (averaged over three synchronized images) shows two spirals rotating in opposite direction captured at $f=70$\,Hz and $\Gamma=23$. The red dash curves highlight the spiral arms and yellow arrows denote the rotation directions. The inset shows the order parameter $I_{\rm RMS}$ vs. $\Gamma$ at $f=80$\,Hz for various time step $\Delta t$ and both directions of $\Gamma$ variations.}
\end{figure}

There is a clear threshold for the patterns to emerge, as the step change of the order parameter $I_{\rm rms}$ with $\Gamma$ in the inset of Fig.~\ref{pd} indicates. $I_{\rm rms}=\sqrt{\sum{(I(x,y)-\bar{I})^2}}$, with $I(x,y)$ the image intensity at the position $(x,y)$ and $\bar{I}$ the spatial average of $I(x,y)$, is chosen as the order parameter. No clear hysteresis could be detected within the error of $\Gamma$, as step increasing ($\Gamma_{\rm inc}$) or decreasing ($\Gamma_{\rm dec}$) the acceleration for various waiting times $\Delta t$ yields the same threshold, which is determined by the averaged steepest slope of $I_{\rm rms}$. Within the range of frequencies investigated, the threshold is increasing monotonically with frequency and insensitive to the geometry of the container and the liquid content $W$: measurements for $f=80$\,Hz with a square shaped container (with an inner side length $10$\,cm) and with a varying $W$ up to 9.4\% yield qualitatively the same spiral patterns and quantitatively the same threshold. The weak dependence on the liquid content, even for the regime where large liquid clusters form, can be attributed to the insensitivity of the rigidity of the wet granular material on $W$ \cite{Scheel08}, which is important for the kink dynamics because the granular layer will be bent and stretched. Nevertheless, wetting is a necessary condition for the present instability because no patterns could be observed without wetting liquid. Below $55$\,Hz, visualization of the pattern is prohibited by the particles sticking to the lid.

In the pattern forming regime, the number of rotating spirals increases with $\Gamma$. The spirals all show period tripling, and may rotate in the same or opposite directions (see the snapshot in Fig.~\ref{pd}). They are considered to be stable, since no decay could be observed for at least $15$ minutes. Even higher $\Gamma$ finally lead to a homogeneous gas-like state, where the particles occupy the whole space of the container. Decreasing $\Gamma$ shows the reversed scenario: from a homogeneous gas state to the spiral patterns, until a flat featureless state appears below the threshold. In contrast to the phase diagram investigated before \cite{Fingerle08,*Huang09}, no granular `gas bubbles' could be observed here, the reason for which is a subject of ongoing investigations.

\begin{figure}
\includegraphics[width = 0.40\textwidth]{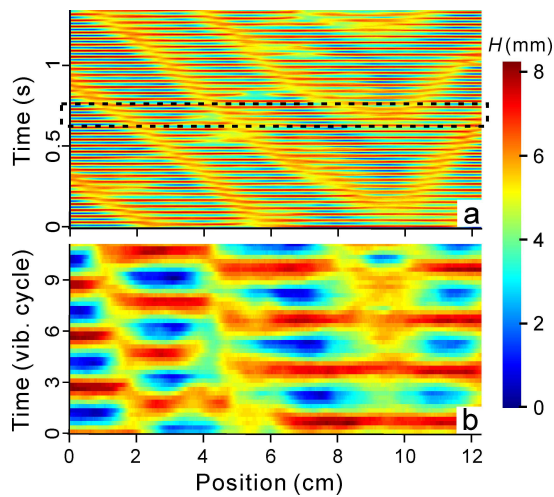} \\ % Here is how to import EPS art
\caption{\label{profile}(color online) Time space plots of the surface profiles obtained with laser profilometry. The long time behavior (a) presents the traveling of the kinks, i.e. spiral arms. The short time behavior (b), corresponding to the region marked with dash line in (a), shows the periodicity of the height fluctuations. Other parameters: $f=80$\,Hz, $\Gamma=32$, frame rate $320$\,Hz, $H$=0 chosen as the lowest surface height.}
\end{figure}

To investigate the spiral dynamics, the surface profiles are measured with laser profilometry. Fig.~\ref{profile} shows the oscillations of an illuminated line close to the core of a rotating spiral arm at different time scales. The long time behavior (Fig.~\ref{profile}(a)) indicates the propagation of the kinks separating regions with different heights $H$. Observations with combined illuminations indicate that the kinks correspond to the arms of spirals. This explains why the arms look brighter: curved surface of the kink regions reflects more light from the LED ring to the camera. The birth of two counter propagating kinks corresponds to the crossing event of one spiral arm with the illuminated line. Since the spiral core is close to the illuminated line, the constant propagation speed suggests a linear relation between the radial distance to the spiral core $r$ and time $t$, i.e. an Archimedean spiral, assuming a constant angular velocity. Once the arms are formed, they are resistant to the disturbances along the propagating arms.

At the time scale of vibration cycles, Fig.~\ref{profile}(b) reveals the period tripling behavior of the height fluctuations, which is in accordance with observations from Fig.~\ref{snapshots}. The fluctuation amplitude being very close to the distance between the surface of the sample at rest and the lid ($\approx7mm$) indicates that the lid plays an important role in developing the spirals. In an open container, no such patterns could be observed.

\begin{figure}
\includegraphics[width = 0.40\textwidth]{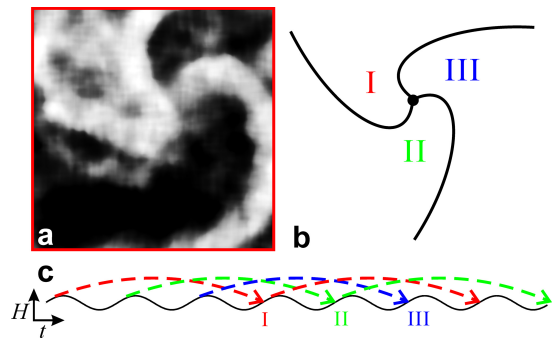}% Here is how to import EPS art
\caption{\label{rotMech}(color online) (a) The covariance between two subsequent images taken as the collision with the container occurs. The brightness corresponds to the mobility of particles. Sketches (b) and (c) indicate the spatiotemporal chirality of the pattern. The black line in (c) indicates the positions of the vibrating bottom, the dash lines correspond to the center of mass in regions I, II and III. $H$ denotes the height.}
\end{figure}

Fig.~\ref{rotMech} is meant to illustrate the rotation mechanism of
the spiral arms. Fig.~\ref{rotMech}(a) shows the spatially resolved covariance between two subsequent images, which is calculated by $\sum_{\Delta x, \Delta y=0}^{d}I_t(x+\Delta x,y+\Delta y) I_{t+1}(x+\Delta x,y+\Delta y)$, where $t$ denotes time and $d$ the particle size in pixels. The time step ($2.5$\,ms) is short enough for the mobility of particles to be captured (bright colors corresponding to smaller covariance, i.e. higher mobility). As interpreted in Fig.~\ref{rotMech}(b) and (c), the higher mobility of particles (granular temperature) in region I arises from the fact that those particles have just undergone a collision with the container bottom. Region II and III will undergo this event in the consequent two vibration cycles. The bright line in Fig.~\ref{rotMech}(a) separating region II and III is due to the rotation of the spiral arm.

The qualitative idea to explain the rotation of the spirals is the assumption that regions with higher granular temperature will expand into `cold' regions. Thus we expect a maximum velocity for the spiral arm separating region I and II because region II is the `coldest' due to the longest time interval having elapsed after the last collision with the bottom. Therefore the predominating motion of the spiral arms will be the rotation towards a region that is going to collide with the container in the next vibration cycle. Reversing the time sequence among region I, II and III -- corresponding to spatiotemporal pattern with different chirality -- would reverse the rotation direction of the spiral. Thus we expect rotating spirals for periodic states with multiples of the vibration period which are larger than 2. The migration fronts shown in Fig.~\ref{snapshots} are explained by the same mechanism: the front moves towards the region lagging behind with a phase shift of $120^{\circ}$, corresponding to a time lapse of one vibration cycle.

To get a clue why period tripling is so dominant in our apparatus, we use a single particle bouncing model to estimate the periodicity, which is the period scaled by $1/f$. It treats the sample as one wet particle colliding inelastically with the container. In addition to model used to describe agitated dry granular layers \cite{Melo95l}, it considers the capillary force acting on the granular layer \cite{Herminghaus05} and collisions with the lid of the container. Initially the particle moves together with the container bottom until the vibrating acceleration is large enough to overcome the capillary and gravitational forces acting on the particle. After the first detachment, the particle takes a parabolic flight until the next collision occurs. After that, it may detach immediately or stay with the container, depending on the colliding phase. This process will run iteratively until the particle rests on the container bottom again, which ends the cycle and determines the periodicity.  Experimentally the periodicity is determined by fitting a Gaussian profile around the peak of the surface height fluctuations in Fourier space.

\begin{figure}
\includegraphics[width = 0.40\textwidth]{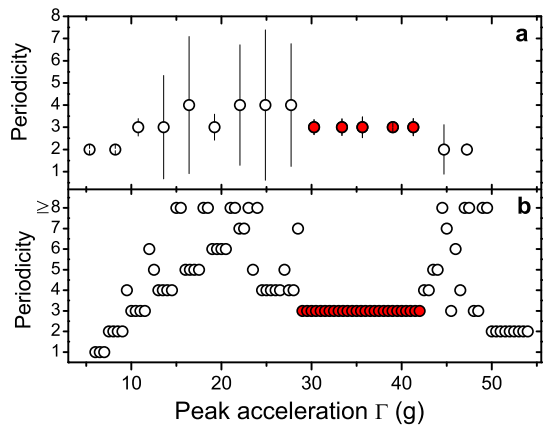}\\
\caption{\label{per}(color online) Periodicity as a function of driving acceleration for $f=80$\,Hz, obtained experimentally from the surface height fluctuations (a), and estimated by a single particle bouncing model (b). Errors in (a) correspond to the variance of the Gaussian profiles fitted to the spectrum of the height fluctuations. Red solid symbols highlight period 3 regimes above the threshold. Other parameters are the same as in Fig.\ref{pd}.}
\end{figure}

Fig.~\ref{per} indicates that the simplified model captures the dependance of periodicity on $\Gamma$ and predicts the threshold for time dependent patterns. Periodic oscillations appear above a certain threshold, where the period is an integer multiple of the driving period. Increasing $\Gamma$ leads to larger periods because the free flying time increases as long as the particles do not touch the lid of the container. Once they touch the lid, the periodicity starts to decay due to the additional impact from the lid. Different from the model, experimental results show high uncertainties of periodicity below $\Gamma \approx 30$, which arise from the intermittent fluctuations with small amplitude. For $\Gamma>30$, both the experiment and the model reveal a wide range with period tripling, which corresponds to the time dependent patterns observed in the experiment. In this regime, the periodicity is robust against small surface fluctuations. Note that the wide period tripling region appears after impact with the lid, which can thus be considered as a technical trick to expand the existing range of spirals.

In conclusion, three armed rotating spirals are found to be the dominant pattern in agitated wet granular matter. The preferred period tripling could be understood by a simplified model considering complete inelastic collisions between the granular layer and the container. The spatiotemporal symmetry breaking connected with the period tripling explains the rotation of the spiral arms. A more detailed model will be needed in order to calculate the exact curvature of the spiral arms. In fact, this pattern could provide a convenient test of more elaborate theories for the dense flow of wet granular matter.

The authors thank K.\ \"Otter for technical support, and F.\ Gollwitzer for help in building the setup. Inspiring discussions with I.\ Aranson, E.\ Meron, F.\ Rietz, K.\ R\"oller, C.\ Kr\"ulle and E.\ Martens are gratefully acknowledged. This work is partly supported by Forschergruppe 608 `Nichtlineare Dynamik komplexer Kontinua'.

%\bibliography{spiral}

\begin{thebibliography}{26}%
\makeatletter
\providecommand \@ifxundefined [1]{%
 \@ifx{#1\undefined}
}%
\providecommand \@ifnum [1]{%
 \ifnum #1\expandafter \@firstoftwo
 \else \expandafter \@secondoftwo
 \fi
}%
\providecommand \@ifx [1]{%
 \ifx #1\expandafter \@firstoftwo
 \else \expandafter \@secondoftwo
 \fi
}%
\providecommand \natexlab [1]{#1}%
\providecommand \enquote  [1]{``#1''}%
\providecommand \bibnamefont  [1]{#1}%
\providecommand \bibfnamefont [1]{#1}%
\providecommand \citenamefont [1]{#1}%
\providecommand \href@noop [0]{\@secondoftwo}%
\providecommand \href [0]{\begingroup \@sanitize@url \@href}%
\providecommand \@href[1]{\@@startlink{#1}\@@href}%
\providecommand \@@href[1]{\endgroup#1\@@endlink}%
\providecommand \@sanitize@url [0]{\catcode `\\12\catcode `\$12\catcode
  `\&12\catcode `\#12\catcode `\^12\catcode `\_12\catcode `\%12\relax}%
\providecommand \@@startlink[1]{}%
\providecommand \@@endlink[0]{}%
\providecommand \url  [0]{\begingroup\@sanitize@url \@url }%
\providecommand \@url [1]{\endgroup\@href {#1}{\urlprefix }}%
\providecommand \urlprefix  [0]{URL }%
\providecommand \Eprint [0]{\href }%
\providecommand \doibase [0]{http://dx.doi.org/}%
\providecommand \selectlanguage [0]{\@gobble}%
\providecommand \bibinfo  [0]{\@secondoftwo}%
\providecommand \bibfield  [0]{\@secondoftwo}%
\providecommand \translation [1]{[#1]}%
\providecommand \BibitemOpen [0]{}%
\providecommand \bibitemStop [0]{}%
\providecommand \bibitemNoStop [0]{.\EOS\space}%
\providecommand \EOS [0]{\spacefactor3000\relax}%
\providecommand \BibitemShut  [1]{\csname bibitem#1\endcsname}%
\let\auto@bib@innerbib\@empty
%</preamble>
\bibitem [{\citenamefont {Ferriani}\ \emph {et~al.}(2008)\citenamefont
  {Ferriani} \emph {et~al.}}]{Ferriani08}%
  \BibitemOpen
  \bibfield  {author} {\bibinfo {author} {\bibfnamefont {P.}~\bibnamefont
  {Ferriani}} \emph {et~al.},\ }\href {\doibase 10.1103/PhysRevLett.101.027201}
  {\bibfield  {journal} {\bibinfo  {journal} {Phys. Rev. Lett.}\ }\textbf
  {\bibinfo {volume} {101}},\ \bibinfo {pages} {027201} (\bibinfo {year}
  {2008})}\BibitemShut {NoStop}%
\bibitem [{spi()}]{spiralzoom}%
  \BibitemOpen
  \href {spiralzoom.com} {}\bibinfo {note}
  {{h}ttp://spiralzoom.com}\BibitemShut {NoStop}%
\bibitem [{\citenamefont {Morris}\ \emph {et~al.}(1993)\citenamefont {Morris}
  \emph {et~al.}}]{Morris93}%
  \BibitemOpen
  \bibfield  {author} {\bibinfo {author} {\bibfnamefont {S.~W.}\ \bibnamefont
  {Morris}} \emph {et~al.},\ }\href {\doibase 10.1103/PhysRevLett.71.2026}
  {\bibfield  {journal} {\bibinfo  {journal} {Phys. Rev. Lett.}\ }\textbf
  {\bibinfo {volume} {71}},\ \bibinfo {pages} {2026} (\bibinfo {year}
  {1993})}\BibitemShut {NoStop}%
\bibitem [{\citenamefont {Plapp}\ \emph {et~al.}(1998)\citenamefont {Plapp}
  \emph {et~al.}}]{Plapp98}%
  \BibitemOpen
  \bibfield  {author} {\bibinfo {author} {\bibfnamefont {B.~B.}\ \bibnamefont
  {Plapp}} \emph {et~al.},\ }\href {\doibase 10.1103/PhysRevLett.81.5334}
  {\bibfield  {journal} {\bibinfo  {journal} {Phys. Rev. Lett.}\ }\textbf
  {\bibinfo {volume} {81}},\ \bibinfo {pages} {5334} (\bibinfo {year}
  {1998})}\BibitemShut {NoStop}%
\bibitem [{\citenamefont {Winfree}(1972)}]{Winfree72}%
  \BibitemOpen
  \bibfield  {author} {\bibinfo {author} {\bibfnamefont {A.~T.}\ \bibnamefont
  {Winfree}},\ }\href@noop {} {\bibfield  {journal} {\bibinfo  {journal}
  {Science}\ }\textbf {\bibinfo {volume} {175}},\ \bibinfo {pages} {634}
  (\bibinfo {year} {1972})}\BibitemShut {NoStop}%
\bibitem [{\citenamefont {Tyson}\ \emph {et~al.}(1988)\citenamefont {Tyson}
  \emph {et~al.}}]{Tyson88}%
  \BibitemOpen
  \bibfield  {author} {\bibinfo {author} {\bibfnamefont {J.~J.}\ \bibnamefont
  {Tyson}} \emph {et~al.},\ }\href@noop {} {\bibfield  {journal} {\bibinfo
  {journal} {Physica D}\ }\textbf {\bibinfo {volume} {32}},\ \bibinfo {pages}
  {327} (\bibinfo {year} {1988})}\BibitemShut {NoStop}%
\bibitem [{\citenamefont {Oliver}\ \emph {et~al.}(1993)\citenamefont {Oliver}
  \emph {et~al.}}]{Steinbock93}%
  \BibitemOpen
  \bibfield  {author} {\bibinfo {author} {\bibfnamefont {S.}~\bibnamefont
  {Oliver}} \emph {et~al.},\ }\href@noop {} {\bibfield  {journal} {\bibinfo
  {journal} {Nature}\ }\textbf {\bibinfo {volume} {366}},\ \bibinfo {pages}
  {322} (\bibinfo {year} {1993})}\BibitemShut {NoStop}%
\bibitem [{\citenamefont {Vanag}\ \emph {et~al.}(2001)\citenamefont {Vanag}
  \emph {et~al.}}]{Vanag01}%
  \BibitemOpen
  \bibfield  {author} {\bibinfo {author} {\bibfnamefont {V.~K.}\ \bibnamefont
  {Vanag}} \emph {et~al.},\ }\href {\doibase 10.1126/science.1064167}
  {\bibfield  {journal} {\bibinfo  {journal} {Science}\ }\textbf {\bibinfo
  {volume} {294}},\ \bibinfo {pages} {835} (\bibinfo {year}
  {2001})}\BibitemShut {NoStop}%
\bibitem [{\citenamefont {Davidenko}\ \emph {et~al.}(1992)\citenamefont
  {Davidenko} \emph {et~al.}}]{Davidenko92}%
  \BibitemOpen
  \bibfield  {author} {\bibinfo {author} {\bibfnamefont {J.}~\bibnamefont
  {Davidenko}} \emph {et~al.},\ }\href@noop {} {\bibfield  {journal} {\bibinfo
  {journal} {Nature}\ }\textbf {\bibinfo {volume} {355}},\ \bibinfo {pages}
  {349} (\bibinfo {year} {1992})}\BibitemShut {NoStop}%
\bibitem [{\citenamefont {Cherry}\ \emph {et~al.}(2008)\citenamefont {Cherry}
  \emph {et~al.}}]{Fenton08}%
  \BibitemOpen
  \bibfield  {author} {\bibinfo {author} {\bibfnamefont {E.~M.}\ \bibnamefont
  {Cherry}} \emph {et~al.},\ }\href@noop {} {\bibfield  {journal} {\bibinfo
  {journal} {New J. Phys.}\ }\textbf {\bibinfo {volume} {10}},\ \bibinfo
  {pages} {125016} (\bibinfo {year} {2008})}\BibitemShut {NoStop}%
\bibitem [{\citenamefont {Rosen}\ and\ \citenamefont
  {Lissauer}(1988)}]{Rosen88}%
  \BibitemOpen
  \bibfield  {author} {\bibinfo {author} {\bibfnamefont {P.~A.}\ \bibnamefont
  {Rosen}}\ and\ \bibinfo {author} {\bibfnamefont {J.~J.}\ \bibnamefont
  {Lissauer}},\ }\href@noop {} {\bibfield  {journal} {\bibinfo  {journal}
  {Science}\ }\textbf {\bibinfo {volume} {241}},\ \bibinfo {pages} {690}
  (\bibinfo {year} {1988})}\BibitemShut {NoStop}%
\bibitem [{Cas()}]{Cassini10}%
  \BibitemOpen
  \href {photojournal.jpl.nasa.gov/catalog/PIA12545} {}\bibinfo {note}
  {{h}ttp://photojournal.jpl.nasa.gov/catalog/PIA12545}\BibitemShut {NoStop}%
\bibitem [{\citenamefont {Spahn}\ and\ \citenamefont
  {Schmidt}(2006)}]{Spahn06}%
  \BibitemOpen
  \bibfield  {author} {\bibinfo {author} {\bibfnamefont {F.}~\bibnamefont
  {Spahn}}\ and\ \bibinfo {author} {\bibfnamefont {J.}~\bibnamefont
  {Schmidt}},\ }\href@noop {} {\bibfield  {journal} {\bibinfo  {journal}
  {GAMM-Mitt.}\ }\textbf {\bibinfo {volume} {29}},\ \bibinfo {pages} {115}
  (\bibinfo {year} {2006})}\BibitemShut {NoStop}%
\bibitem [{\citenamefont {Goldhirsch}(2003)}]{Goldhirsch03}%
  \BibitemOpen
  \bibfield  {author} {\bibinfo {author} {\bibfnamefont {I.}~\bibnamefont
  {Goldhirsch}},\ }\href@noop {} {\bibfield  {journal} {\bibinfo  {journal}
  {Annu. Rev. Fluid Mech.}\ }\textbf {\bibinfo {volume} {35}},\ \bibinfo
  {pages} {267} (\bibinfo {year} {2003})}\BibitemShut {NoStop}%
\bibitem [{\citenamefont {Aranson}\ and\ \citenamefont
  {Tsimring}(2006)}]{AransonRMPl}%
  \BibitemOpen
  \bibfield  {author} {\bibinfo {author} {\bibfnamefont {I.~S.}\ \bibnamefont
  {Aranson}}\ and\ \bibinfo {author} {\bibfnamefont {L.~S.}\ \bibnamefont
  {Tsimring}},\ }\href {\doibase 10.1103/RevModPhys.78.641} {\bibfield
  {journal} {\bibinfo  {journal} {Rev. Mod. Phys.}\ }\textbf {\bibinfo {volume}
  {78}},\ \bibinfo {eid} {641} (\bibinfo {year} {2006})}\BibitemShut {NoStop}%
\bibitem [{\citenamefont {Jop}\ \emph {et~al.}(2006)\citenamefont {Jop} \emph
  {et~al.}}]{Pouliquen06}%
  \BibitemOpen
  \bibfield  {author} {\bibinfo {author} {\bibfnamefont {P.}~\bibnamefont
  {Jop}} \emph {et~al.},\ }\href@noop {} {\bibfield  {journal} {\bibinfo
  {journal} {Nature}\ }\textbf {\bibinfo {volume} {441}},\ \bibinfo {pages}
  {727} (\bibinfo {year} {2006})}\BibitemShut {NoStop}%
\bibitem [{\citenamefont {Jiang}\ and\ \citenamefont {Liu}(2007)}]{Liu07l}%
  \BibitemOpen
  \bibfield  {author} {\bibinfo {author} {\bibfnamefont {Y.}~\bibnamefont
  {Jiang}}\ and\ \bibinfo {author} {\bibfnamefont {M.}~\bibnamefont {Liu}},\
  }\href {\doibase 10.1103/PhysRevLett.99.105501} {\bibfield  {journal}
  {\bibinfo  {journal} {Phys. Rev. Lett.}\ }\textbf {\bibinfo {volume} {99}},\
  \bibinfo {pages} {105501} (\bibinfo {year} {2007})}\BibitemShut {NoStop}%
\bibitem [{\citenamefont {Bocquet}\ \emph {et~al.}(2002)\citenamefont {Bocquet}
  \emph {et~al.}}]{Bocquet02}%
  \BibitemOpen
  \bibfield  {author} {\bibinfo {author} {\bibfnamefont {L.}~\bibnamefont
  {Bocquet}} \emph {et~al.},\ }\href@noop {} {\bibfield  {journal} {\bibinfo
  {journal} {C. R. Physique}\ }\textbf {\bibinfo {volume} {3}},\ \bibinfo
  {pages} {207} (\bibinfo {year} {2002})}\BibitemShut {NoStop}%
\bibitem [{\citenamefont {Herminghaus}(2005)}]{Herminghaus05}%
  \BibitemOpen
  \bibfield  {author} {\bibinfo {author} {\bibfnamefont {S.}~\bibnamefont
  {Herminghaus}},\ }\href {\doibase 10.1080/00018730500167855} {\bibfield
  {journal} {\bibinfo  {journal} {Adv. Phys.}\ }\textbf {\bibinfo {volume}
  {54}},\ \bibinfo {pages} {221} (\bibinfo {year} {2005})}\BibitemShut
  {NoStop}%
\bibitem [{\citenamefont {de~Bruyn}\ \emph {et~al.}(2001)\citenamefont
  {de~Bruyn}, \citenamefont {Lewis}, \citenamefont {Shattuck},\ and\
  \citenamefont {Swinney}}]{Bruyn01l}%
  \BibitemOpen
  \bibfield  {author} {\bibinfo {author} {\bibfnamefont {J.~R.}\ \bibnamefont
  {de~Bruyn}}, \bibinfo {author} {\bibfnamefont {B.~C.}\ \bibnamefont {Lewis}},
  \bibinfo {author} {\bibfnamefont {M.~D.}\ \bibnamefont {Shattuck}}, \ and\
  \bibinfo {author} {\bibfnamefont {H.~L.}\ \bibnamefont {Swinney}},\ }\href
  {\doibase 10.1103/PhysRevE.63.041305} {\bibfield  {journal} {\bibinfo
  {journal} {Phys. Rev. E}\ }\textbf {\bibinfo {volume} {63}},\ \bibinfo
  {pages} {041305} (\bibinfo {year} {2001})}\BibitemShut {NoStop}%
\bibitem [{\citenamefont {Efsen}\ \emph {et~al.}(1995)\citenamefont {Efsen}
  \emph {et~al.}}]{Raton95s}%
  \BibitemOpen
  \bibfield  {author} {\bibinfo {author} {\bibfnamefont {J.}~\bibnamefont
  {Efsen}} \emph {et~al.},\ }\enquote {\bibinfo {title} {Handbook of
  non-invasive methods and the skin},}\ \ (\bibinfo  {publisher} {CRC Press
  Inc., Boca Raton},\ \bibinfo {year} {1995})\ Chap.~\bibinfo {chapter} {8},
  p.~\bibinfo {pages} {97}\BibitemShut {NoStop}%
\bibitem [{\citenamefont {Faraday}(1831)}]{Faraday1830}%
  \BibitemOpen
  \bibfield  {author} {\bibinfo {author} {\bibfnamefont {M.}~\bibnamefont
  {Faraday}},\ }\href@noop {} {\bibfield  {journal} {\bibinfo  {journal}
  {Philos. Trans. R. Soc. London}\ }\textbf {\bibinfo {volume} {121}},\
  \bibinfo {pages} {299} (\bibinfo {year} {1831})}\BibitemShut {NoStop}%
\bibitem [{\citenamefont {Melo}\ \emph {et~al.}(1995)\citenamefont {Melo},
  \citenamefont {Umbanhowar},\ and\ \citenamefont {Swinney}}]{Melo95l}%
  \BibitemOpen
  \bibfield  {author} {\bibinfo {author} {\bibfnamefont {F.}~\bibnamefont
  {Melo}}, \bibinfo {author} {\bibfnamefont {P.~B.}\ \bibnamefont
  {Umbanhowar}}, \ and\ \bibinfo {author} {\bibfnamefont {H.~L.}\ \bibnamefont
  {Swinney}},\ }\href {\doibase 10.1103/PhysRevLett.75.3838} {\bibfield
  {journal} {\bibinfo  {journal} {Phys. Rev. Lett.}\ }\textbf {\bibinfo
  {volume} {75}},\ \bibinfo {pages} {3838} (\bibinfo {year}
  {1995})}\BibitemShut {NoStop}%
\bibitem [{\citenamefont {Scheel}\ \emph {et~al.}(2008)\citenamefont {Scheel}
  \emph {et~al.}}]{Scheel08}%
  \BibitemOpen
  \bibfield  {author} {\bibinfo {author} {\bibfnamefont {M.}~\bibnamefont
  {Scheel}} \emph {et~al.},\ }\href {\doibase 10.1038/nmat2117} {\bibfield
  {journal} {\bibinfo  {journal} {Nature Mater.}\ }\textbf {\bibinfo {volume}
  {7}},\ \bibinfo {pages} {189} (\bibinfo {year} {2008})}\BibitemShut {NoStop}%
\bibitem [{\citenamefont {Fingerle}\ \emph {et~al.}(2008)\citenamefont
  {Fingerle} \emph {et~al.}}]{Fingerle08}%
  \BibitemOpen
  \bibfield  {author} {\bibinfo {author} {\bibfnamefont {A.}~\bibnamefont
  {Fingerle}} \emph {et~al.},\ }\href@noop {} {\bibfield  {journal} {\bibinfo
  {journal} {New J. Phys.}\ }\textbf {\bibinfo {volume} {10}},\ \bibinfo
  {pages} {053020} (\bibinfo {year} {2008})}\BibitemShut {NoStop}%
\bibitem [{\citenamefont {Huang}\ \emph {et~al.}(2009)\citenamefont {Huang}
  \emph {et~al.}}]{Huang09}%
  \BibitemOpen
  \bibfield  {author} {\bibinfo {author} {\bibfnamefont {K.}~\bibnamefont
  {Huang}} \emph {et~al.},\ }\href@noop {} {\bibfield  {journal} {\bibinfo
  {journal} {Eur. Phys. J-Spec. Top.}\ }\textbf {\bibinfo {volume} {179}},\
  \bibinfo {pages} {25} (\bibinfo {year} {2009})}\BibitemShut {NoStop}%
\end{thebibliography}

\end{document}